\documentclass[journal]{IEEEtran}
\IEEEoverridecommandlockouts                              

\usepackage{color}
\usepackage{silence}
\usepackage{cite}
\usepackage{algorithmic}
\usepackage{algorithm}
\usepackage{amssymb}
\usepackage{amsmath}
\usepackage{multirow}
\usepackage{multicol}   
\usepackage[justification=centering]{caption}
\usepackage{amsmath}
\usepackage{footnote}
\makesavenoteenv{tabular}
\usepackage{graphicx}

\usepackage{caption}
\usepackage{subcaption}

\usepackage{epstopdf,float,amsfonts}
\usepackage{epsfig}

\title{Scalable Voltage Control using Structure-Driven  Hierarchical Deep Reinforcement Learning }
\author{Sayak Mukherjee,~\IEEEmembership{Member, IEEE,}
           Renke Huang,~\IEEEmembership{Senior Member,~IEEE,} \\ Qiuhua Huang,~\IEEEmembership{ Member,~IEEE},  
         Thanh Long Vu,~\IEEEmembership{ Member,~IEEE}, and Tianzhixi Yin
\thanks{Pacific Northwest National Laboratory (PNNL) is operated by Battelle for the U.S. Department of Energy (DOE) under Contract DE-AC05-76RL01830. This work was funded by DOE ARPA-E OPEN 2018 Program.
\textit{(Corresponding author: Renke Huang, Qiuhua Huang})}
\thanks{S. Mukherjee, R. Huang, Q. Huang, T. L. Vu, and T. Yin are with the Pacific Northwest National Laboratory, Richland, WA, USA.
Emails: \{\texttt{sayak.mukherjee, renke.huang, qiuhua.huang, thanhlong.vu, tianzhixi.yin}\}@\texttt{pnnl.gov.}}}

\begin{document}

\maketitle
 \begin{abstract}
     This paper presents a novel hierarchical deep reinforcement learning (DRL) based design for the voltage control of power grids. DRL agents are trained for fast, and adaptive selection of control actions such that the voltage recovery criterion can be met following disturbances. Existing  voltage control techniques suffer from the issues of speed of operation, optimal coordination between different locations, and scalability. We exploit the area-wise division structure of the power system to propose a hierarchical DRL design that can be scaled to the larger grid models. We employ an enhanced augmented random search algorithm that is tailored for the voltage control problem in a two-level architecture. We train area-wise decentralized RL agents to compute lower-level policies for the individual areas, and concurrently train a higher-level DRL agent that uses the updates of the lower-level policies to efficiently coordinate the control actions taken by the lower-level agents.  Numerical experiments on the IEEE benchmark $39-$bus model with $3$ areas demonstrate the advantages and various intricacies of the proposed hierarchical approach.  
 \end{abstract}
\vspace{-.35 cm}
\section{Introduction}
Resiliency and security of bulk power system is of vital importance with the significant increase of penetration of inverter-interfaced resources, and dynamic loads. Fault induced delayed voltage recovery (FIDVR) \cite{fidvr} causes bus voltage magnitudes to stay at a significantly
reduced levels for several seconds after fault clearing, due to the stalling of induction motor loads and prolonged tripping \cite{fidvr_induction}, which could lead to wide-area outages \cite{AEMO2017report}. One of the industry standard practices to counteract voltage instabilities is to perform load shedding \cite{van2007vs_overview}. Standard approaches such as empirical rule based approach using pre-specified thresholds \cite{Lefebvre2003ls4HQ} without coordination \textcolor{black}{usually result in unnecessary load shedding to safeguard the system.} On the other hand, security constrained power flow \cite{Misra2017} or model predictive control (MPC) \cite{Glavic2006MPC} based approaches would require accurate knowledge about the model of the system along with the requirement of computationally expensive real-time optimization solutions. 

Recently, \textcolor{black}{to reduce the reliance on accurate model}, data-driven approaches with adaptive characteristics are being investigated, for example, decision tree based algorithm \cite{Genc2010_DT}, a hierarchical extreme-learning machine based algorithm  \cite{Qiao2020hierarchical_LS}, etc. \color{black} Reinforcement learning (RL) based data-driven approaches can directly optimize load shedding actions using output and reward feedback from the the grid \cite{huang2019loadshedding_DRL, zhang2018loadshedding_DRL, huang2020accelerated}. RL marries the utilities of adaptive \cite{adaptive} and optimal control \cite{optimal_book}, where the RL agent interacts with the system, and optimizes its actions to accumulate maximum possible reward over episodes. There are varieties of Markov decision process (MDP)-based RL algorithms  using value-based or policy gradient based or a combination of these approaches in works such as  \cite{mnih2013atari,LillicrapHPHETS15,schulman2017ppo,schulman2015trpo,SAC}, to name a few.
In our previous works, we have designed a deep Q-learning based \cite{huang2019loadshedding_DRL}, and subsequently, an accelerated deep RL (DRL) approach in \cite{huang2020accelerated} for voltage control. Along these lines of research, this paper proposes a scalable DRL algorithm that exploits the structure of the power grid and can be efficiently applied to large-scale power system models. \color{black}
\par

\textcolor{black}{It is worth to note that most of the existing RL-based emergency grid control approaches are trained and deployed in a centralized manner, where the controller observes the measurements at multiple buses and sends the control signals to several load buses across the grid to perform emergency load shedding. This centralized paradigm requires long communication links from the control center to several remote locations, and a failure in, or a cyberattack to,  the control center or a link can make the control performance ineffective. In addition, training the centralized RL on the large-scale power grids would require very long, if not unacceptable, training times to train excessive number of neural network parameters}. Otherwise, if the policy network is trained with insufficient number of parameters then the control performance may degrade. \textcolor{black}{An alternative solution is to train and implement the RL control in a completely local manner, in which each local RL control in a pre-specified geographical area is designed to recover the local bus voltages if fault happens in that area. However, similar to the rule-based under voltage load shedding, the lack of coordination among the local RL controls will lead to unnecessary load shedding, and can also lead to insufficient recovery in other areas}. 

\textcolor{black}{
To alleviate these concerns of the centralized and local RL framework, in this paper we propose an innovative hierarchical RL framework for the training and deployment of intelligent emergency load shedding.} Remarkably, our design exploits the \textit{structural properties} of the power grid to incorporate modularity in both training and implementation, and can provide sufficient control performance on par or better than centralized policies. Architecturally, we employ a two-level RL policy where the lower level policies are trained based on the clustering or area division structure of the grid independently avoiding the \textit{non-stationarity} issue \cite{marl1_review} in mutli-agent RL training, and the higher level agent coordinates them to achieve sufficient control performance. 
\textcolor{black}{To save the training time, we will train the lower-level RLs in a parallel manner using multiple cores, and their policy updates are used to train the higher-level RL concurrently. In addition, as we use multiple fault rollouts for the lower level policies across the grid, the training of higher-level RL only takes a limited number of faults used in each of lower-level RL training. This training framework significantly reduces the simulation time in higher-level RL training, while still ensuring the coordination among lower-level RLs and offering robust performance to multiple contingencies.} 
Our approach is different than classical hierarchical RL techniques \cite{Feudal, HIRO}  as we utilize the structural clustering of the grid to enforce hierarchy, rather than temporal abstractions. On the algorithmic side, we propose the distributed hierarchical learning employing an enhanced version of augmented random search methodology \cite{mania2018_ARS}, namely parallel augmented random search (PARS) \cite{huang2020accelerated} that explores the parameter space of policies and shown to be of better computational efficiency than the other model-free methods used for multi-agent training such as \cite{marl1_review, marl4_maddpg}. 
\par
\textit{Contributions:} The main contributions of the paper are:

\begin{itemize}
    \item We present a novel structure-driven, hierarchical, multi-agent DRL algorithm for emergency voltage control design that can be scaled to larger power system models with faster learning and increase in the modularity. We exploit the inherent area divisions of the grid, and propose a \textit{structure-exploiting} DRL design by incorporating few traits of hierarchical and multi-agent learning to propose a two-level architecture. 
    
    \item We employ two concurrent training mechanisms. On one computing branche, we train the area-wise decentralized policies based on the fault scenarios in the corresponding areas. On the other branch, we simultaneously train a higher-level RL agent to intelligently supervise and coordinate the lower-level agents, thereby saving considerable training time 
     compared to centralized training. 
    \item We demonstrate the performance of our approach with IEEE $39-$bus power system model with $3$ areas. Our approach can be easily extendable to larger industrial models paving the way for real-world implementation of artificial intelligence (AI) driven decision making for wide-area voltage control. \textcolor{black}{We show that our proposed hierarchical RL can save 60$\%$ training time in comparison to the centralized full-scale training, while achieving comparable, if not better, control policies.}
\end{itemize} \color{black}
\par
\textit{Organization:} The rest of the paper is organized as follows. The formulation of solving dynamic voltage stability control problem via RL methods is described in Section II. Section III describes our proposed structure-driven hierarchical RL design. Test results are shown in Section IV, and concluding remarks are given in Section V.
\section{Dynamic Voltage Stability Control Problem}

We start by describing the the power system dynamics in a differential-algebraic form as:
\vspace{-.12 cm}
\begin{align}\label{sys}
    &\dot{x} = f(x,x_a,u), \; x(0)=x_0, x_a(0)=x_{a_0},\\
    &g(x,x_a,u)=0, \label{pf}
\end{align}
where the grid dynamic states are denoted as $x \in \mathbb{R}^{n_x}$, the algebraic variables are denoted as $x_a \in \mathbb{R}^{n_{x_a}}$, and $u \in \mathbb{R}^{n_u}$ denotes the controls. The non-linear function $f: \mathbb{R}^{n_x} \times \mathbb{R}^{n_{x_a}} \times \mathbb{R}^{n_u} \to \mathbb{R}^{n_x}$ captures the dynamics of the grid states, and \eqref{pf} with $g(.) = 0$ characterizes the non-linear power flow. In the optimal control setting, the objective would be to minimize cost functionals such as $\int_{0}^{\infty}J(x(t), x_a(t), u(t))dt$. The RL controllers can optimize from experience by repeatedly interacting with the environment such as complex grid simulators without knowing the dynamic models.

\par
The RL problem is conventionally formulated in the Markov decision processs (MDP) framework, which requires a fully or partially observable state space $\mathcal{X}$, action space $\mathcal{U}$, the transition dynamics $\mathcal{T} : \mathcal{X} \times \mathcal{U} \to \mathcal{X}$ to find $x_{t+1} \in \mathcal{X}$ based on the current state $x_t \in \mathcal{X}$, and action $u_t \in \mathcal{U}$, and a scalar reward $r \in \mathbb{R}$ associated with the action and the state transition. For the dynamic voltage stability control problem we define the complete MDP as follows.
\begin{itemize}
    \item \textit{Observation space:} Accessing all the dynamic states of the power system is a difficult task, and the operators can only measure a limited number of algebraic states and  outputs. For the voltage control problem, the bus voltage magnitudes $V(t)$ and the remaining percentage of the loads at the controlled buses $P_{D}(t)$ are easily measurable, and therefore, considered in the the observation space denoted as $\mathcal{Y} \subset \mathcal{X} \cup \mathcal{P}$, $\mathcal{X} \subseteq \mathbb{R}^{n_x}, $where $\mathcal{P} \subseteq \mathbb{R}^{n_{x_a}}$ denotes the space pertaining to the algebraic variables. The observation variables $y_t \in \mathcal{Y}$ are continuous in nature.
    \item \textit{Action space:} We consider controllable loads as actuators where shedding locations are generally set by the utilities by solving a rule-based optimization problem for secure grid operation. We consider the operator can shed upto $20\%$ of the total load at a particular bus at any given time instant. The action space is continuous with $[-0.2,\;0]$ range where $-0.2$ denotes the $20\%$ load shedding, with the actions denoted as $u_t \in \mathcal{U} \subseteq \mathbb{R}^{n_u}$. 
    \item \textit{Policy class:} Policies are the mappings from $y_t$ to the actions $u_t$ denoted as $\pi(.): y_t \to u_t $. The learning design will compute this policy optimally such that it can achieve the desired objective. In our learning design, we consider the long short term memory (LSTM) network\cite{Schmidhuber1997_LSTM} due to its capability of automatically learning to capture the temporal dependence over multiple time steps. 
    \item \textit{Transition dynamics:} The dynamics of the bus voltage magnitudes and remaining percentage of the loads are governed by the differential-algebraic equations (DAEs) as in \eqref{sys}-\eqref{pf}. Following a disturbance input $d(t)$, the transition is captured by the flow map $F(.)$ such as,
    \begin{align}
        y(t+1)= F(y(t), u(t),d(t)), y(t_0)=y_0.
    \end{align}
    Please note in our RL algorithm, we will not require the knowledge about this transition dynamics and only use the measurements of the trajectories of $y(t)$, and $u(t)$ to design the control. The framework is flexible, and also accounts for stochastic transitions which are important for power systems with increased uncertainties. 
    \item \textit{Rewards:} The scalar rewards $r \in \mathbb{R}$ are designed to achieve the voltage control objective, i.e, to keep the voltages of all the buses within the safe recovery profile as given in Fig. 1. During the hierarchical RL control design, we will describe in detail about the reward definitions for different layers of RL agents. 
\end{itemize}
\begin{figure}[t]
    \centering
    \includegraphics[width = .8\linewidth]{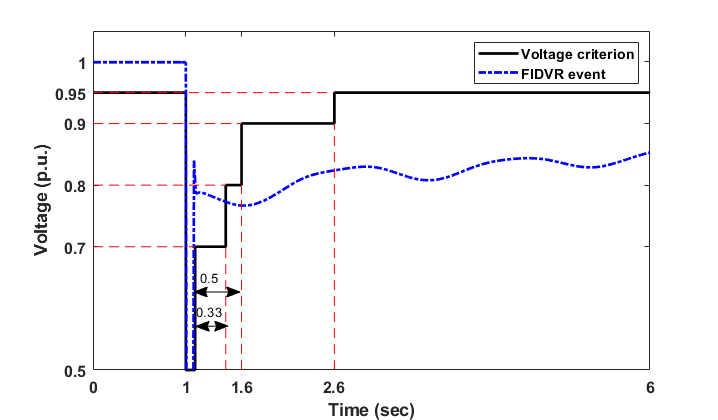}
    \caption{\small{Safe voltage recovery criterion}}
    \label{fig:tvrc}
    \vspace{-.4 cm}
\end{figure}
\vspace{-.4 cm}
\section{Hierarchical reinforcement Learning}
Performing full-scale centralized RL control for large-scale practical grid models will definitely encounter the curse of dimensionality. This leads to the idea of hierarchy by exploiting the interconnected nature of the grid and relatively localized effect of the voltage stability problem. We exploit the area divisions where different control areas are often equipped with localized controls and sometimes managed by different utilities, and interconnected via tie-lines. Our approach is composed of two concurrent stages. We train lower-level DRL policies for individual areas, and parallely train a higher-level coordinator policy using the policy updates from the lower-level training. The higher-level agent activates the lower-level policies in a coordinated manner to achieve system level voltage control objective. The coupled lower and higher levels of training helps in bringing scalability to the design by parallelizing the computing burden to each individual decentralized agents and their coordinator. Fig. \ref{fig:scheme} shows a conceptual overview of our approach.
\vspace{-.4 cm}
\subsection{Learning Area-wise Decentralized Policies:}
We consider the grid to be divided into $r$ non-overlapping zones with their corresponding bus indices enumerated in the set $\mathcal{I}_i, i=1,2,..,r$. The bus voltage magnitudes  in these areas are denoted by $V_i \in \mathbb{R}^{n_i}, i={1,2,..,r},$ and the remaining percentage of loads at their controlled buses by $P_{Di} \in \mathbb{R}^{m_i}, i={1,2,..,r}.$ We denote the set of buses with controllable loads for the areas by $a_1, \dots, a_r$. The lower-level policy for area-$i$ is being trained with multiple faults at area-$i$. The neighbor buses for each individual areas, denoted as $\mathcal{N}_{Vi}$, are those buses where the impact of faults from area $i$ is considerable. Therefore, for the area $i$, the observations $y_i$ are consisting of $y_i = \{ V_i, (V_j, j\in \mathcal{N}_{Vi}), P_{Di}\}$, for $i=1,\dots,r$, and the load shedding control actions are denoted by $u_i$ where $i \in a_i$, for $i=1,\dots,r$. Please note that although we take neighboring feedbacks, we use the term \textit{decentralized} to signify that the lower policies are independently trained only for their corresponding areas.  
\\
\subsubsection{ Finding local neighborhood of the $i^{th}$ area}  
We start by exploring area $i$ by running contingencies at the control actuation locations to emulate an impulse-like input that can excite all the inherent oscillation modes. We use the safe voltage recovery profile $\sigma(t)$ as shown in Fig. 1 to set the criterion of neighbor buses following contingencies. We first check if any of the bus voltage magnitudes of the area $j, j \neq i$ violates the voltage safety profile $\sigma(t)$ for a contingency in area $i$. An area is not considered as neighbors if none of the neighbor buses violate $\sigma(t)$(or a very small number of violations). Else, we select representative buses that violates the safety profile as $\mathcal{N}_{Vi} = \{ k, k \in  \mathcal{I}_{j\neq i} | \; V_{jk}(t) - \sigma(t) < 0 , t \geq T_{pf}\}$, $T_{pf}$ is the fault clearing time. When multiple buses in area $j$ violate the safety profile, representative worst-affected buses are selected, where we use the criterion of bus voltage magnitude nadirs to be lower than $0.75$ p.u. Minimal set of feedbacks also decreases the number of weights to be trained in the policy network. 
\begin{figure}[t]
    \centering
    \includegraphics[width = .95\linewidth, trim = 4 4 4 4, clip]{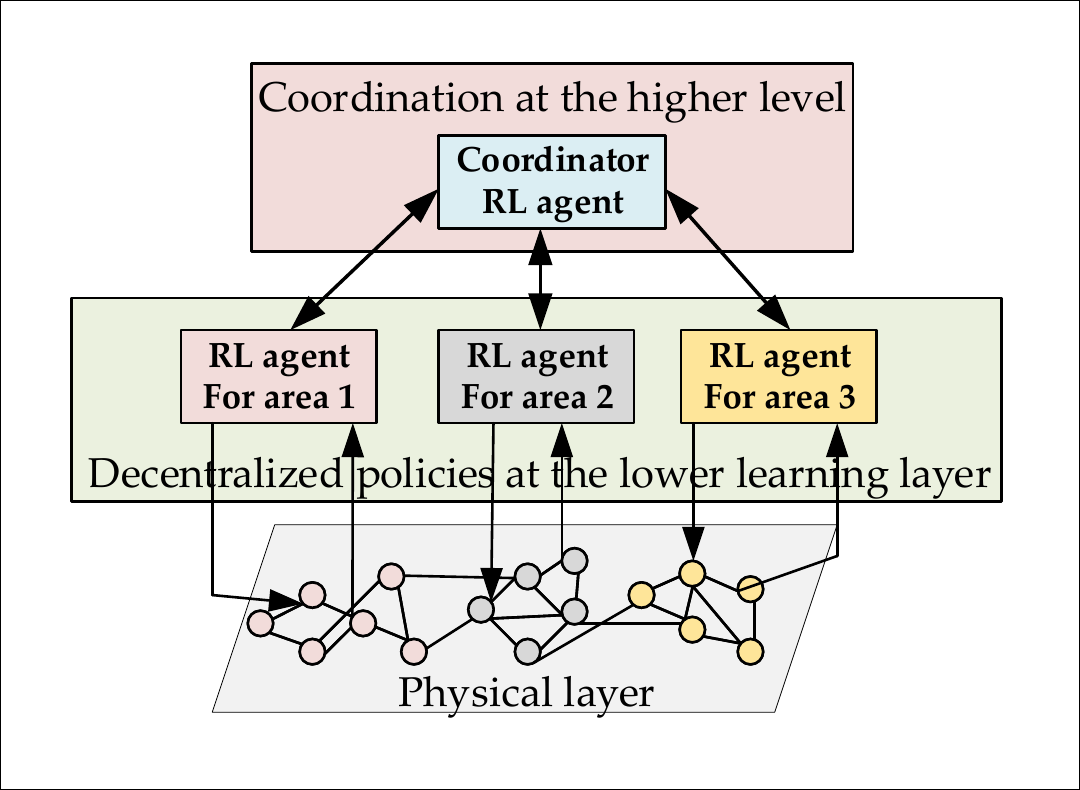}
    \caption{\small{Schematic of the Hierarchical RL architecture}}
    \label{fig:scheme}
    \vspace{-.56 cm}
\end{figure}
\subsubsection{Decentralized Augmented Random Search}
The decentralized policies are denoted as $\pi_{\theta_i}, i=1,2,\dots,r$, where $\theta_i$ denotes the policy parameters, which in our design denotes the LSTM weights and biases for the lower-level policy of the $i^{th}$ area. To train a sufficiently good policy, we specify a reward function for the respective areas. Let us denote the bus voltage magnitudes in the observations of area $i$ be $V_{ai} = \{ V_i, (V_j, j\in \mathcal{N}_{Vi})\}$. \color{black} RL agent's objective is defined as to maximize the expected reward, where
the reward $r(t)$ at time $t$ for area $i$ is set as follows:
\begin{align}
\label{eq.reward}
r_i(t) = 
\begin{cases}
&-1000  {\;\;\; \emph{if}\;\;\;} V_{aij}(t)<0.95,\;\;\; T_{pf} +4<t \\
& c_1 \sum_j \Delta V_{aij}(t) - c_2 \sum_j \Delta P_{Lij} (p.u.) - c_3 v_{ivld}, \\&{\;\;\; \emph{\mbox{otherwise, where}},}
\end{cases}
\end{align}
\vspace{-.5 cm}
\footnotesize
\begin{align}
\Delta V_{aij}(t) =
\begin{cases}
&\min \{V_{aij}(t) - 0.7, 0\},  \emph{\;if\;} t \in (T_{pf}, T_{pf} +0.33) \\
&\min \{V_{aij}(t) - 0.8, 0\}, \emph{\;if\;} t \in (T_{pf} +0.33, T_{pf} +0.5 ) \\
&\min \{V_{aij}(t) - 0.9, 0\} , \emph{\;if\;}  t \in (T_{pf} +0.5, T_{pf} +1.5) \\
&\min \{V_{aij}(t) - 0.95, 0\} , \emph{\;if\;} t \in t > T_{pf} +1.5.
\end{cases}\nonumber
\end{align}
\normalsize

In the reward function \eqref{eq.reward}, $T_{pf}$ is the time instant of fault clearance;
$V_{aij}(t)$ is the voltage magnitude for bus $j$ correspond to area $i$;  $\Delta P_{Lij} (t)$ is the
load shedding amount in p.u. at time step $t$ for load bus $j$ corresponding to area $i$; 
invalid action penalty $v_{ivld}$ if the DRL agent still provides
load shedding action when the load at a specific bus has already been fully shed at the previous time step when
the system is within normal operation. $c_1 , c_2$, and $c_3$ are the weight factors for the above three parts.

\color{black}
Algorithm $1$ presents the steps to compute the decentralized policies for the individual areas. Please note that each area contains a set of fault buses $l_i, i=1,\dots,r$, (each with cardinality $|l_i|$) within an area for $p_i, i=1,\dots,r,$ different fault duration. Step $8$ describes the rollouts or trajectory simulations with these faults applied in the emulator. We will describe the steps shortly after presenting the coordinator design.  
\begin{algorithm}[]
\caption{Decentralized ARS (DARS) - Lower Level}
\begin{algorithmic}
\STATE 1. \texttt{Set} hyper-parameters for the areas: For the $i^{th}$ area, select the step size $\alpha_i$, the number of policy perturbation
directions per iteration $N_i$, standard deviation of the
exploration noise $\nu_i$, the number of top-performing perturbed
directions selected for updating weights $b_i$, the number of
rollouts per perturbation direction $m_i, (m_i = |l_i|\cdot p_i)$ and the decay rate $\epsilon_i$, for all $i=1,\dots,r$.

\FOR{$i =1,\dots,r$} 
\STATE 2. \texttt{Perform} faults at the controlled buses, and store measurements of bus voltage magnitudes.
\STATE 3. \texttt{Select} the neighborhood of the area $i$ ($\mathcal{N}_{Vi}$) using the safe voltage recovery profile $\sigma(t)$.
\STATE 4. \texttt{Select} representative bus voltage magnitudes from the neighbors to construct the observations $y_i \in \mathbb{R}^{\bar{n}_i}$.
\STATE 5. \texttt{Initialize} the policy weights $\theta_i \in \mathbb{R}^{n_{\theta_i}}$ randomly with
the running mean of observation states $\mu_0 = \mathbf{0} \in \mathbb{R}^{\bar{n}_i}$,
and the running standard deviation of the observation states $\Sigma_0 = \mathbf{I}_{\bar{n}}$, and the total iteration number $H$.
\WHILE{ iteration $k \leq H$}
\STATE 6. \texttt{Sample} $N_i$ number of random directions $\delta_{i1},\dots,\delta_{iN_i} \in \mathbb{R}^{n_{\theta_i}}$.
\FOR{each $\delta_{ij}, j=1,\dots, N_i$}
\STATE 7. \texttt{Add} perturbations to policy weights $\theta_{ijk+} =\theta_{i(k-1)} + \nu_i \delta_{ij}, \theta_{ijk-} =\theta_{i(k-1)} - \nu_i \delta_{ij}$.
\STATE 8. \texttt{Perform} $2m_i$ rollouts according to the selected faults in the area $i$ corresponding to the policy weights $\theta_{ijk+}$, and $\theta_{ijk-}$, respectively, and then compute average rewards $r_{ijk+}, r_{ijk-}$, for $j=1,\dots,N_i$.
\STATE 9. \texttt{Evaluate} trajectories: During the rollout, the observations of area $i$ at time-step $t$ $y_{ik,t}$ are normalized and then passed through the policy $\pi_{ik}(.)$ to generate the action $u_{ik,t}$, and applied to the environment:
\begin{align}
    &y_{ik,t} = \frac{y_{ik,t} - \mu_{i(k-1)}}{\Sigma_{i(k-1)}},\\
    & u_{ik,t} = \pi_{ik}(y_{ik,t}),\\
    & y_{ik,t+1} = F(y_{ik,t},u_{ik,t},d_i,t),
\end{align}
where $d_i,t$ is the disturbance in the $i^{th}$ area. The running mean and the standard deviations are also updated.
\ENDFOR
\STATE 10. \texttt{Sort} the directions $\delta_{ij}$  based on $\mbox{max}(r_{ijk+},r_{ijk-}), j=1,2,..,N_i$. Select top $b_i$ directions, and calculate their standard deviations $\sigma_{bi}$.
\STATE 11. \texttt{Update} the policy weight of the $i^{th}$ ARS learner, $i=1,2..,r$:
\begin{align}
    \theta_{i(k+1)} = \theta_{ik} + \frac{\alpha_i}{b_i  \sigma_{bi}} \sum_{d=1}^{b_i} (r_{idk+} - r_{idk-})\delta_{id}
\end{align}
\STATE 12. \texttt{Decay} the step size and standard deviation by the rate $\epsilon_i$.\\  
\ENDWHILE
\ENDFOR
\RETURN 13. Decentralized lower level policies $\pi_{\theta_i}$, $i=1,2,\dots,r$. 
\end{algorithmic}
\end{algorithm}
\vspace{-.4 cm}
\subsection{Concurrent Learning of Coordinator Policy:}
To this end, we have the mechanism of generating locally learned policies $\pi_{\theta_i}$ for the areas $i=1,\dots,r$. If we perform a fully decentralized design, then one can just use this lower level policies based on the fault location, i.e., if a fault occurs in area $i$, then only the controller at the area $i$ will be actuated. However, 
that may not sufficient to improve the dynamic performance in the neighboring areas. Moreover, without intelligent coordination, all the lower-layer agents will be activated and although that may produce sufficient performance on voltages, it will lead to expensive inefficient load shedding. Motivated from these concerns, we will now design a \textit{higher-level coordinator} that can synchronize the lower-level actions efficiently. Mathematically, this means the higher-level coordinator policy $\pi^c$ needs to intelligently select lower level policies $\pi_{\theta_i}$'s depending on the location and severity of the faults. The Coordinating RL agent will have discrete action spaces that selects different areas. We now describe the mathematical framework and the algorithm.  


\subsubsection{Mathematical formalization of the transition dynamics}
When one of the lower-level RL agents are actuated, for example the one in the area $i$, then the power system transition dynamics is given as:
\begin{align}\label{tran_decen1}
    y(t+1) = F(y(t), \pi_{\theta_i}(y_i (t)), d_i (t)).
\end{align}
If we consider noisy dynamic behaviour then we have,
\begin{align}\label{tran_decen2}
    y_{t+1} \sim P(y_{t+1}\;| y_{t}, \pi_{\theta_i}, d_i),
\end{align}
where the argument $d_i$ signifies that the distribution $P(.)$ is subjected to a fault at the area $i$ at the buses given in the set $l_i$. Once we design the coordinator, which we will describe shortly, then the coordinator will select the higher level policies $\pi^c$, and that in turn will activate lower level policies $\pi_{\theta_i}$ in the grid. If multiple RL agents are actuated, then the resultant closed-loop dynamics will be due to their joint action. We refer this idea of selecting one or multiple lower-level agents by the higher-level coordinator as the \textit{field of vision} of the coordinator at each time step with the notation $\pi^c \to \pi_{\theta} \in \{\pi_{\theta_i}, i=1,\dots,r \}$.  Therefore, the power system dynamics due to the combined action of the coordinator and decentralized agents is given by:
\begin{align}\label{tran_combined1}
    y(t+1) = F(y(t), \pi^c \to \pi_{\theta}, d_{j \in [1:r]}(t)),
\end{align}
or the stochastic equivalent
\begin{align}\label{tran_combined2}
    y_{t+1} \sim P(y_{t+1}\;| y_{t}, \pi^c \to \pi_{\theta}, d_{j \in [1:r]}(t)).
\end{align}
Here the disturbance $d_j(t)$ can occur in any part of the grid.
Next, we look into the action and the observation spaces of the coordinator. 
\setlength{\textfloatsep}{0pt}

\begin{algorithm}[t]
\caption{Coordinated ARS (CARS) - Higher Level}
\begin{algorithmic}
\STATE 1. \texttt{Set} the hyper-parameters  $\alpha^c, N^c, \nu^c, b^c, m^c ( = |\cup_{i=1}^{r} c_i|\cdot p^c),$ and $\epsilon^c$ following similar descriptions as in Alg. 1, and the positive integers $H_l, H_c$, and \texttt{initialize} the  coordinator policy parameters $\theta^c$. 
\FOR{iteration $k=1,\dots,H$}
\STATE 2. \texttt{Sample} randomly $N$ perturbation directions $\delta^c_1,..,\delta^c_{N^c}$ which are of the same dimension as $\theta^c$.
\FOR{ each direction $\delta^c_j, j=1,\dots,N^c$}
\STATE 3. \texttt{Add} positive and negative perturbations to $\theta^c$: $\theta^{c}_{k+} = \theta^c_{k-1} + \nu^c \delta^c_{i}, \theta^c_{k-} = \theta^c_{(k-1)} - \nu^c \delta^c_{i}$.
\STATE 4. \texttt{Use} lower layer policies $\pi_{\theta_i}, i=1,\dots,r$ from DARS:   \\
\textbf{if} $\pi_{\theta_i}, i=1,\dots,r,$ are not converged:\\
A. \texttt{Read} $\pi_{\theta_i}, i=1,\dots,r,$ once each time after running an interval of $H_l$ number of iterations in DARS.\\
B. \texttt{Implement} the updated policies  $\pi_{\theta_i}, i=1,\dots,r,$ for the subsequent $H_c$ iterations in CARS.\\
\textbf{else}:\\
C. \texttt{Implement} the converged policies  $\pi_{\theta_i}, i=1,\dots,r,$\\
\textbf{endif}\\
\STATE 5. \texttt{Run} similar to Steps 8 and 9 as in Alg. 1 to perform $2m^c$ fault rollouts, and evaluate trajectories following Consideration 2 in III.B.5.
\ENDFOR
\STATE 6. \texttt{Run} similar to Steps 10, 11, and 12 as in Alg. 1 to perform the policy update:
\begin{align}
    \theta^c_{(k+1)} = \theta^c_{k} + \frac{\alpha^c}{b^c \sigma^c_b} \sum_{d=1}^{b^c} (r_{kd+} - r_ {kd-})\delta^c_{d}.
\end{align}
\ENDFOR
\RETURN 7. Coordinator policy $\theta^c$.
\end{algorithmic}
\end{algorithm}
\subsubsection{Action space of coordinator} The actions that the higher level coordinator can take are of discrete values. Based on the fault location, the action space can be made restricted. This is because if a fault occurs in area $i$, we expect to activate the controls in an intelligent way for only the areas that are the physical neighbors of area $i$. Let us denote the physical neighbor areas by $\mathbb{N}_i$. Please note that $\mathbb{N}_i$ is different than the neighbors  $\mathcal{N}_{Vi}$ that we have computed in the Step 1 of DARS. For example, in the IEEE 39-bus system, the physical neighbors of area $1$ are $\mathbb{N}_1 = [2,3]$, and discrete action space when the fault occurs in area $1$ is given as $\mathcal{U}^c_1 = [1, (1,2), (1,3), (1,2,3)]$. Similarly action spaces for other areas $\mathcal{U}^c_i$ can be designed. For a particular fault, the actions can be restricted to $\pi^c_i \in \mathcal{U}^c_i$. However, when starting with an unrestricted action space of the coordinator, the reward design will help the RL agent to select the optimal set of lower-level actuators to gain high rewards. That will make the action selection fully automated and the operator does not need to pass any information about the location of the fault.
\subsubsection{Observation space of coordinator} The observation space of the coordinator is composed of the the minimum of the instantaneous bus voltage magnitudes for individual areas. The instantaneous minimum captures the worst-case behaviour following a fault. For area $i$, the observation used for the learning design is $ \mbox{min}(V_{i}(t))$, where $V_{i}(t)$ are the set of voltage magnitudes of all the buses in the area $i$. We denote the observation space for the coordinator as $y^c = \{y^c_1,\dots, y^c_r \}, y^c_i = \mbox{min}(V_{i}(t)). $
\subsubsection{Reward function of the coordinator} To design the reward function, we use the transient voltage recovery criterion based on the ideal recovery profile $\sigma(t)$.
Then the reward function for the coordinator can be written as:
\begin{align}
    r^c_t &= -1000, \;\; \text{if} \;\;\;\;  \mbox{min}(V_i(t)) < 0.95 p.u., t> T_{pf} + 4,\\
    &= \sum_i c_iq_i - c_2 \sum_j \Delta P_{Lj} (p.u.) - c_3 v_{ivld},, t>T_{pf},\\
    q_i &= \mbox{min} (-(\sigma(t) - y^c_i(t)),0).
\end{align}
\subsubsection{Training} Algorithm 2 gives the steps for the coordinator training. We make the following design considerations to make the overall algorithm scalable and efficient. 

\begin{itemize}
\item {\textit{Consideration 1:}} The higher-level coordinator reads the policies that are being learned in the lower layers via DARS and then implements them in the power system during its roll-outs. However, to achieve steady convergence, the higher-level coordinator reads the lower level policies once after a specified number of iteration interval (say $H_l$) in the lower layers, and then keeps those policies fixed for another pre-specified set of its own iterations (say $H_c$) as in Step 5 of Alg. 2. The CARS iteration starts after running DARS for $H_l$ number of iterations to receive the first set of lower-level policies. When the lower-level policies converge, the corresponding converged policies of the areas are used for subsequent iterations. As the lower layer policies are trained parallelly in a decentralized framework, the grid model does not encounter the issue of non-stationarity during DARS training. However, as these policies may need to be implemented jointly depending on the severity and location of the faults, any approximation during DARS can be compensated by the coordinator training to ensure global performance improvement.

\item {\textit{Consideration 2:}} The computation burden of the coordinator training due to roll-outs are kept similar to that of the lower-level policies. We do that by restricting the upper-level coordinator to train with a limited number of representative faults randomly selected from each of the individual areas. We denote the set of the selected fault buses from the area $i$ by $c_i$ with $|c_i|$ is sufficiently smaller than $ |l_i|$. Therefore, the coordinator performs rollouts at the buses from $\cup_{i=1}^{r} c_i$ along with $p^c$ different fault duration for each of them.  As the CARS algorithm is implemented concurrently with DARS using the updates from the lower layer, the total training time is dictated by that of the higher-level agent. 
\end{itemize}
\subsection{Summary of Algorithmic Steps:}
For both of Algs. 1 and 2, the ARS learner delegates tasks to lower-level workers, gathers rewards, and accordingly updates the policy weights $\theta_i$ and $\theta_c$. The learner communicates with multiple workers which conduct  perturbations (random search) of the policy weights as in Step 7 of Alg. 1 and Step 3 of Alg. 2, and subsequently run fault rollouts for each of the perturbed policies. In an another layer of parallelism, the workers utilize a number of subordinate actors running in parallel where each actor is only responsible for one fault rollout corresponding to the perturbed policy sent by its up-level worker as in Step 8 of Alg. 1 and Step 5 of Alg. 2. In Step 10 of Alg. 1 and in Step 6 of Alg. 2, the ARS learner aggregates the rewards of multiple rollouts conducted by each of the sub-level workers, sorts the directions according to the obtained rewards, and selects the  best-performing directions. Then, in Step 11 of Alg. 1 and in Step 6 of Alg. 2, the ARS learner updates the policy weights $\theta_i$ and $\theta_c$ based on the perturbation results from the top performing workers. Steps 2-4 in Alg. 1 are distinct from the coordinator training where the neighbor buses of individual areas are computed following III.A.1. On the other hand, Steps 4 and 5 in Alg. 2 are designed based on considerations 1 and 2, respectively, as discussed in III.B.5.  
\vspace{-.3 cm}
\section{Test Results}
\begin{figure}[H]
\centering
    \includegraphics[width = .7\linewidth, height = 4.6 cm]{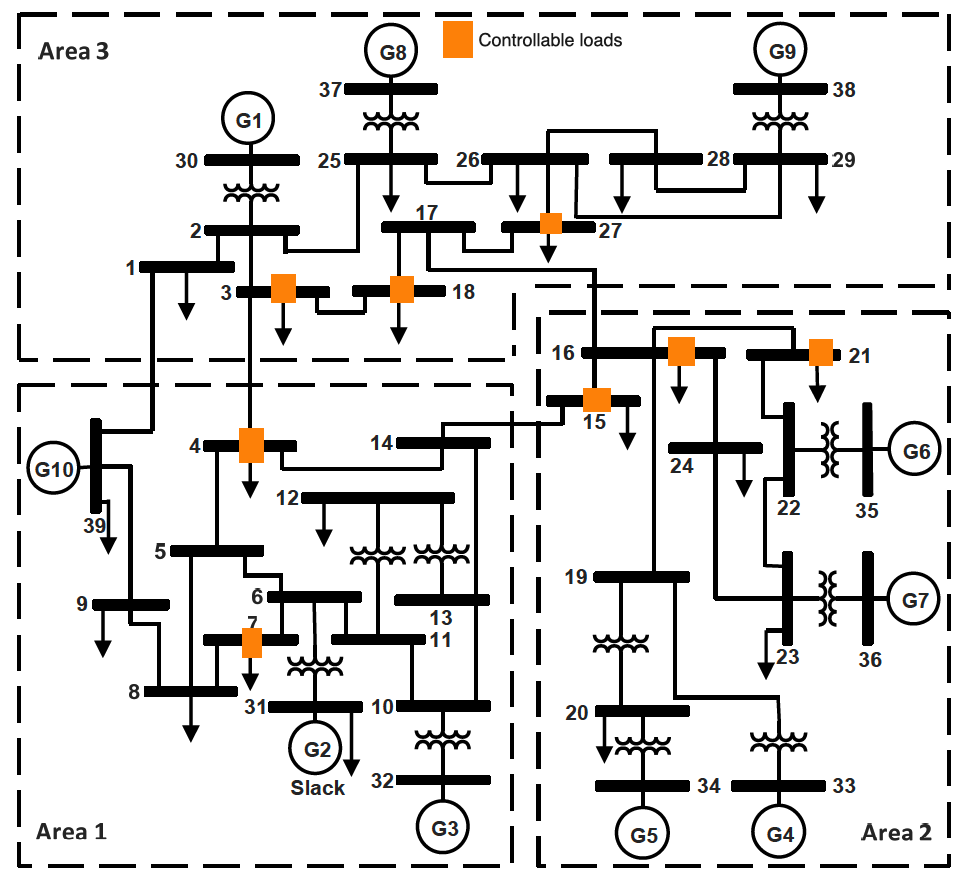}
    \caption{\small{IEEE 39-bus system with three areas}}
    \label{fig:39bus}
    \vspace{-.4 cm}
\end{figure}

We consider the IEEE benchmark $39-$bus, $10-$generator model as our simulation testcase. The proof-of-concept experiments have been performed in a Linux server with AMD Opteron $6272$ processor with $8$ cores per socket and the maximum of $32$ CPUs running at $2.1$ GHz, and $64$ GB memory. The power system simulator is implemented in the GridPack\footnote{https://www.gridpack.org}, and the hierarchical deep RL algorithm is implemented in the python platform. A software setup has been built such that the grid simulations in the GridPack  and the RL iterations in the python can communicate. Fig. \ref{fig:39bus} shows that the grid model can be divided into three distinct areas. In each of the areas, we have considered few buses where controllable loads are connected. For area $1$, the controllable loads are at buses $4$ and $7$, for area $2$, dynamic loads are connected at buses $15, 16$ and $21$, and for area $3$, we have control actions available at buses $3, 18$ and $27$. \par 

\begin{figure*}
    \centering
    \begin{minipage}{0.3\linewidth}
    \centering
    \includegraphics[width = \linewidth]{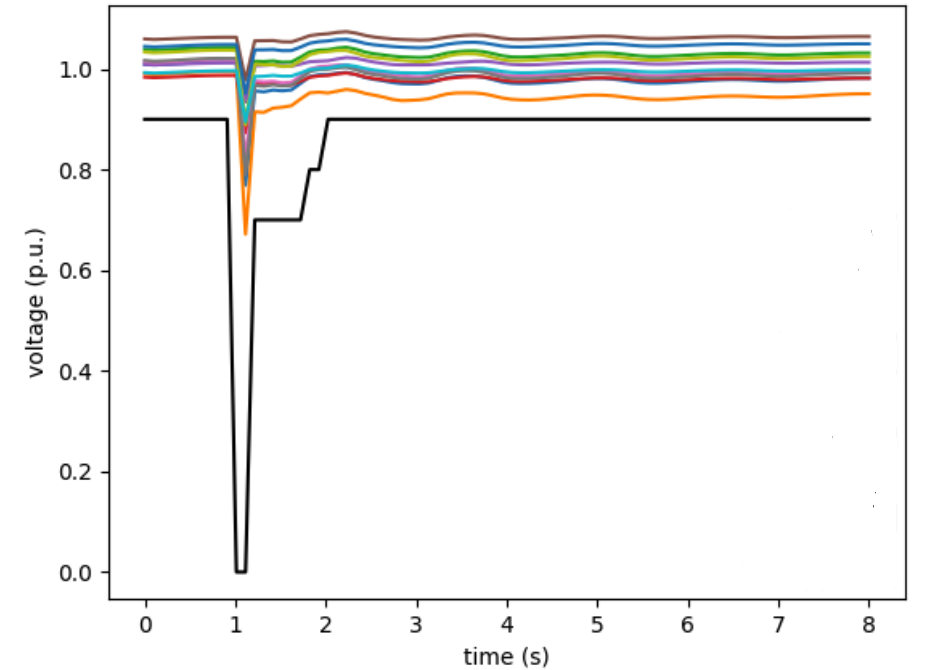}
    \caption{\small{Fault at bus 7, observations at area 2 without any controller}}
    \label{fig:neighbors4}
    \end{minipage}
    \begin{minipage}{0.3\linewidth}
    \includegraphics[width = \linewidth]{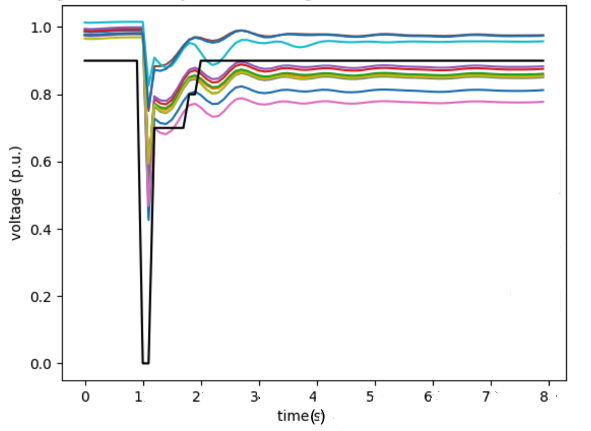}
     \caption{\small{Fault at bus 16, observations at area 1 without any controller}}
    \label{fig:neighbors16}
    \end{minipage}
    \begin{minipage}{0.3\linewidth}
        \centering
        \includegraphics[width = \linewidth]{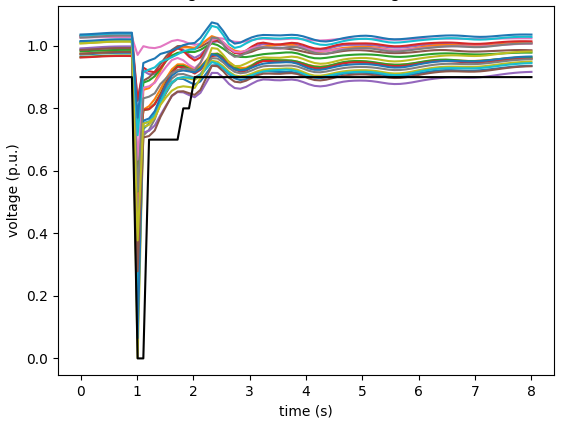} 
        \caption{\small{Insufficient performance of the non-hierarchical centralized policy with fault at bus 16}}
    \label{fig:perfbus16centralized16}
    \end{minipage}
    \begin{minipage}{0.3\linewidth}
        \centering
        \includegraphics[width = \linewidth]{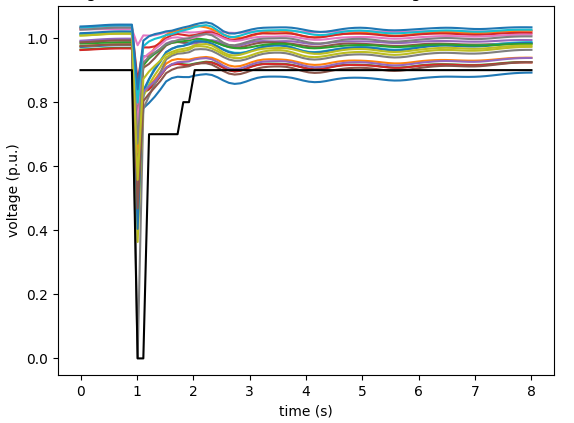} 
        \caption{\small{Insufficient performance of the non-hierarchical decentralized area 2 policy with fault at bus 15}}
    \label{fig:onlyarea2}
    \end{minipage}
    \begin{minipage}{0.3\linewidth}
        \centering
        \includegraphics[width = \linewidth]{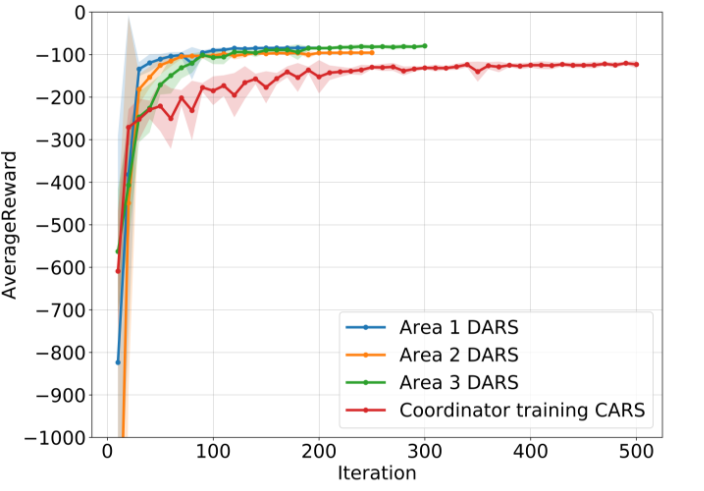} 
        \caption{\small{Training local policies via DARS and coordinator via CARS (shown with multiple seeds)}}
        \label{fig:training_comb}
    \end{minipage}
    \begin{minipage}{0.3\linewidth}
        \centering
        \includegraphics[width = \linewidth]{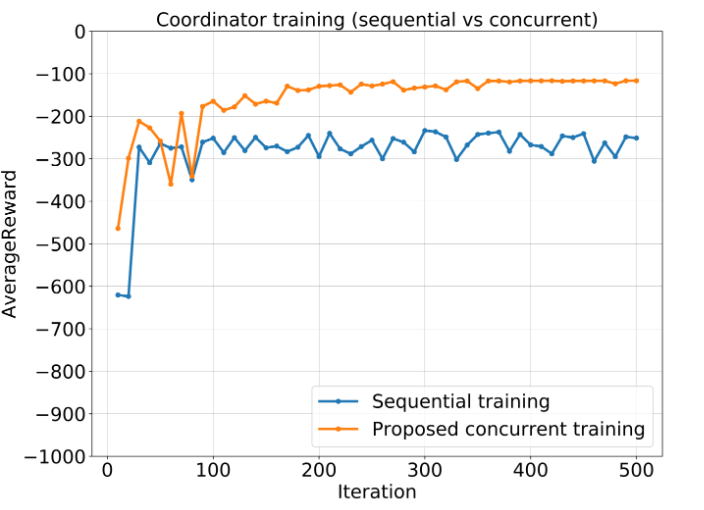} 
        \caption{\small{Coordinator policy convergence with concurrent training vs. sequential training}}
        \label{fig:comp_seq_con}
    \end{minipage}
    \centering
     \centering
     \quad
    \quad
    \begin{minipage}{0.3\linewidth}
        \centering
        \includegraphics[width = \linewidth]{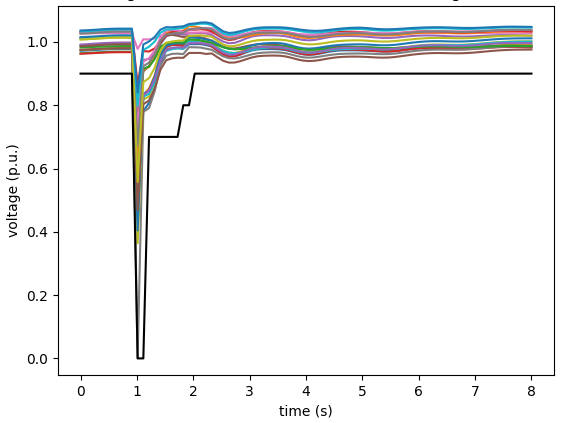} 
        \caption{\small{Performance of the hierarchical RL design with fault at bus 15}}
    \label{fig:perfbus15}
    \end{minipage}
    \quad
    \quad
    \centering
    \begin{minipage}{0.3\linewidth}
        \centering        \includegraphics[width = \linewidth]{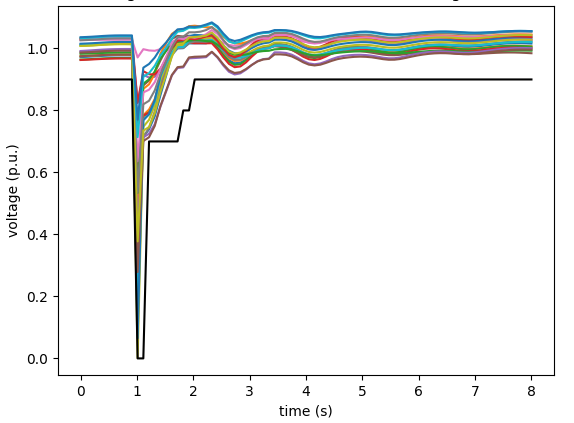} 
        \caption{\small{Performance of the hierarchical RL design with fault at bus 16}}
    \label{fig:perfbus16}
    \end{minipage}
    \begin{minipage}{0.3\linewidth}
        \centering
        \includegraphics[width = \linewidth]{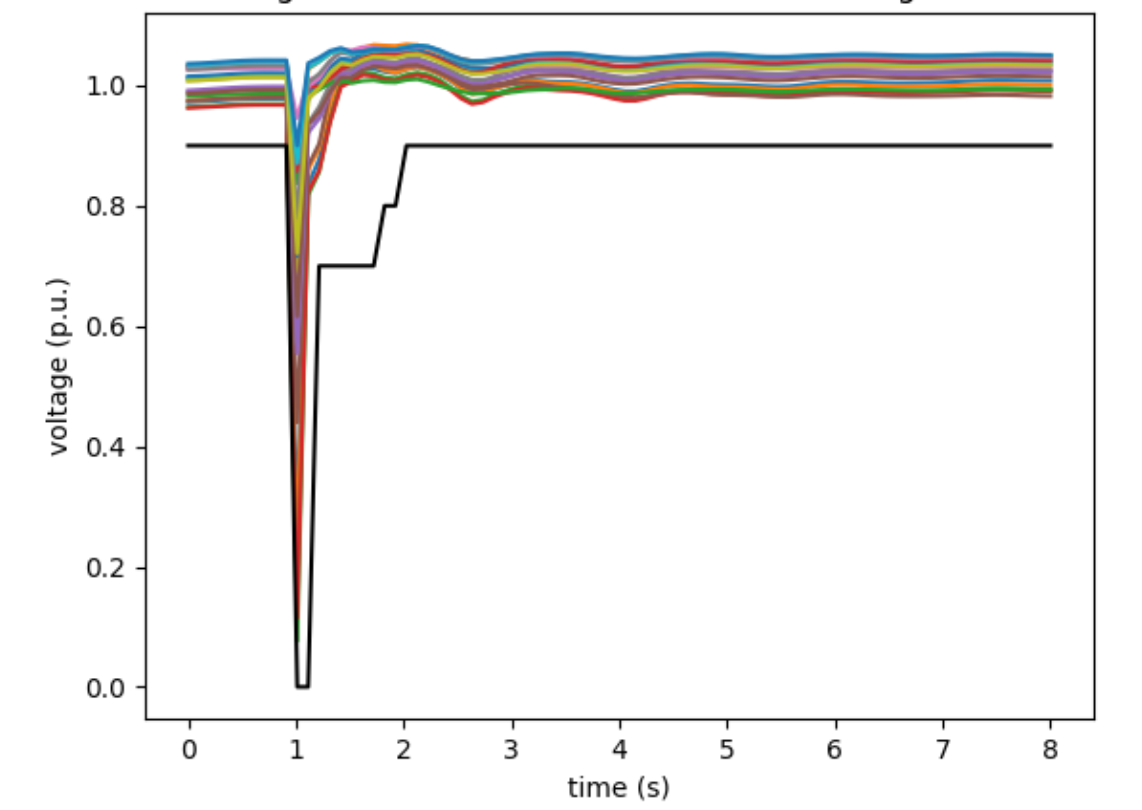} 
        \caption{\small{Performance of the hierarchical RL design with the fault at bus 6}}
    \label{fig:perfbus2}
    \end{minipage}
   \vspace{-0 cm}
\end{figure*}

Before showing the efficacy of the hierarchical design, we first perform two scenarios: a full-scale centralized training, and training only area-wise decentralized policies. For the centralized full-scale design, we consider the policy network with $16$ hidden units in the LSTM layer, and $16$ units in the fully connected layer. Thereafter, we run a centralized augmented random search with faults at buses $4, 12, 11, 24, 19, 20, 1, 25, 26$ with three different duration with $0$ (no fault), $0.05,$ and $0.08$ seconds of faults. Once the centralized ARS training following the steps in \cite{huang2020accelerated} has been performed, we test the learned policy with various different faults in the grid. Fig. \ref{fig:perfbus16centralized16} shows a scenario where the non-hierarchical centralized policy was not able to recover the voltages of all the buses due to fault at bus $16$. This experiment shows that we need to train a larger policy network with more parameters to achieve sufficient performance for the non-hierarchical design. This creates a  bottleneck for large-scale grid models. Moreover, the time taken for full-scale centralized training is $5.705$ hours. We will later show the scalability of our hierarchical algorithm in terms of training time along with using modular policy networks. Considering the second scenario, when we do not implement the higher level coordinator, and only implement the non-hierarchical decentralized policies, Fig. \ref{fig:onlyarea2} shows an example where the control actions at area 2 cannot improve the dynamic behaviour of the grid even if the fault occurred in area 2 itself. These studies show the difficulties in designing a full-scale centralized policy for large-scale grids, and insufficiency of only area-wise decentralized policies without any coordinator.

\begin{table}[t]
\caption{Parameters used in DARS and CARS}
\centering
\begin{tabular}{|l|l|l|l|l|}
\hline
Parameters                                                                         & Area 1       & Area 2       & Area 3       & Coordinator  \\ \hline
\begin{tabular}[c]{@{}l@{}}Policy network size\\ (hidden layers)\end{tabular}      & {[}16, 16{]} & {[}16, 16{]} & {[}16, 16{]} & {[}16, 16{]} \\ \hline
Number of directions                                                               & 14           & 16           & 14           & 20           \\ \hline
Top directions                                                                     & 7            & 8            & 7            & 11           \\ \hline
Maximum iterations                                                                 & 200          & 250          & 300          & 500          \\ \hline
Step size                                                                          & 1            & 1            & 1            & 1            \\ \hline
\begin{tabular}[c]{@{}l@{}}Standard deviation of \\ exploration noise\end{tabular} & 2            & 2            & 2            & 2            \\ \hline
Decay rate                                                                         & 0.995        & 0.995        & 0.995        & 0.995        \\ \hline
\end{tabular}
\label{param}
\end{table}
To this end, we move toward the hierarchical deep RL design. The lower layer policy networks and the coordinator networks are designed with $16$ hidden units in the LSTM layer, and $16$ units in the fully connected layer, and try to learn sufficiently optimized policies. 
We train decentralized policies for the individual areas following DARS and concurrently run CARS to compute optimized coordinator policy. In order to find the neighbor buses for the areas, we follow the procedure described in III.A.1. We excite the grid model with faults at the controlled buses, and observe the bus voltage magnitudes in the neighboring areas. For area 1, we found that the neighbor buses in areas 2 and 3 do not violate the transient voltage recovery profile and we neglect small number of violations, however, during exploration in areas 2 and 3, we found violations in the neighbors. Figs. \ref{fig:neighbors4}-\ref{fig:neighbors16} show two examples, where in Fig. \ref{fig:neighbors4} we found that the impact of fault at bus $7$ on the buses at area 2 does not create violations, however, when fault occurs at bus $16$ in area 2, multiple bus voltage magnitudes in the neighbors violate the safe recovery profile as shown in Fig. \ref{fig:neighbors16}. For area 1, we do not consider any neighbor buses in the observations, however for area 2 and 3 we have the neighbor buses as $\mathcal{N}_{V2} = \{4, 14, 5, 7, 17, 18, 27, 3, 30 \}$, $\mathcal{N}_{V3} = \{4, 14, 5, 7, 16, 15, 21, 24 \}$. For each area, we consider the load buses as the self-observation buses in our design. Thereafter, we implement the area-wise decentralized ARS training for the individual areas as given in the Algorithm 1. The training parameters used for the algorithm is given in Table \ref{param}.
\begin{table}[b]
\vspace{.2 cm}
\caption{Comparing training times of the full-scale centralized and hierarchical approaches}
\centering
\begin{tabular}{|l|l|}
\hline
Centralized full-scale training & 5.705 hours                                                                  \\ \hline
Hierarchical RL training        & \begin{tabular}[c]{@{}l@{}}2.29 hours\\ ($\approx 60\%$ saving)\end{tabular} \\ \hline
\end{tabular}
\label{comp_time}
\end{table}
\begin{figure}[t]
\centering
    \includegraphics[width = .8\linewidth, height = 4.6 cm]{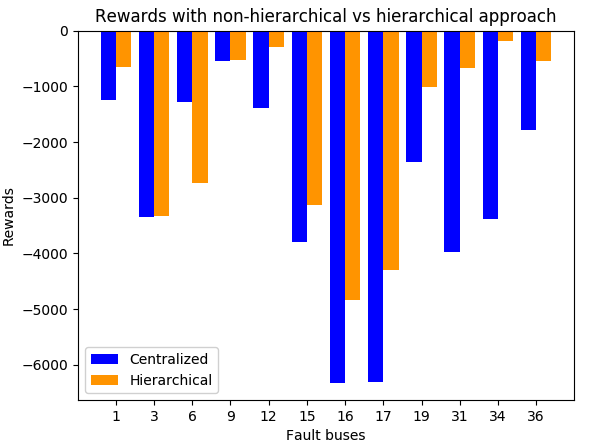}
    \captionsetup{justification=justified}
    \caption{\small{Comparing the rewards shows that the hierarchical design can obtain on par or better performance than the centralized approach. For few critical fault buses such as $15$, hierarchical approach can bring the voltages within the safety profile which the centralized design fails as shown in Figs. 6, 10. When the hierarchical design achieves comparable or lower rewards than the centralized approach, the voltages still remain within the safety bounds such as fault at bus $6$ as shown in Fig. 12}}
    \label{fig:rewards}
\end{figure}
Fig. \ref{fig:training_comb} shows the training profile where average rewards at the end of $10$ iterations are plotted over all the fault rollouts. We run simulations with multiple random seeds to ensure reproducibility of our results. The area-wise decentralized training constituted three separate training for the policies of the individual areas. The training time separately for area 1, 2, and 3 were $35.87$ minutes, $44.04$ minutes, and $53.06$ minutes. Each of these policies are trained with nine fault scenarios within the area. Simultaneously, the lower level policy updates are used as per Algorithm 2 where we train a higher level coordinating RL agent that selects a lower level policy intelligently based on the impact of the fault in the grid. The coordinator randomly uses one fault bus from a set of training buses for each of the areas to generate rollouts. We consider fault duration of $0$ (no fault), $0.05,$ and $0.08$ seconds, totalling $9$ fault scenarios for the CARS training. Fig. \ref{fig:training_comb} shows the training of the coordinator agent following CARS. We use $H_l = 10, H_c =10$. The coordinator training using CARS took $2.26$ hours. As the coordinator iteration starts after initial $10$ iterations of the DARS, and then runs concurrently with the decentralized training, the total training time in this design turns out to be $2.26 + 0.03 = 2.29$ hours. A comparison with the non-hierarchical design is shown in Table \ref{comp_time}, which shows the scalability in computation time for this design with respect to that of the full-scale centralized approach. We also compare the training convergence of the coordinator for the proposed approach to another scenario where the coordinator is trained using the converged lower-level policies in Fig. \ref{fig:comp_seq_con}. This shows that when we pass the run-time lower level updates to the higher level, thereby sending the information on how the lower policies are getting optimized, the coordinator can perform better in terms of convergence and achieve higher rewards. We, then, implement the control architecture as shown in Fig \ref{fig:scheme}, and test with multiple faults with $0.1$s duration at the different bus locations of the grid. Figs. \ref{fig:perfbus15}-\ref{fig:perfbus16} show how hierarchical design can coordinate between the RL agents of different areas to produce much better and desired performance compared to the full-scale centralized design as in Fig. \ref{fig:perfbus16centralized16}, or the decentralized one-area policy as in Fig.  \ref{fig:onlyarea2}. Similar performance is found to be achieved for other fault bus locations as well. Please note that we tested with fault buses that are not used in training to show the robustness and the generalizability of the trained policy. The sufficiently optimized performance of the learned hierarchical policies are also validated in the accumulated rewards over the fault rollouts as shown in Fig. \ref{fig:rewards} with a detailed discussion.   
 \vspace{-.23 cm}
\section{Conclusions and Future Research}
This paper presents a novel scalable reinforcement learning based automated emergency load shedding based voltage control. We exploit the inter-connected structure of the grid to develop a hierarchical control architecture where the lower level RL agents are trained in an area-wise decentralized way, and the higher-level agent is being trained to intelligently coordinate the load shedding actions taken by the lower-level agents. Training of the coordinator and the lower level agents were performed in a concurrent way. The design can compute sufficiently optimized policies in a considerable lower time than the centralized full-scale design by incorporating modularity in training. We have ensured that although the area-wise decentralized policies may not be independently sufficient to recover all the bus voltages after faults, the higher-level coordinating agent can effectively control them to ensure voltage stability under emergency conditions. 
In our future research, we will experiment with our approach for a larger industrial model that can alleviate the insufficiency of the centralized learning approaches.
We will continue to investigate the development of efficient design variants related to multi-agent and hierarchical approaches. 
\bibliographystyle{IEEEtran}
\vspace{-.4 cm}
\bibliography{HDRL4LS}
\end{document}